\documentclass[aps,prl,reprint]{revtex4-1}

\usepackage{graphicx}
\usepackage{dcolumn}
\usepackage{amsfonts}
\usepackage{amsmath}

\usepackage{fancyhdr}

\begin{document}

\title{Coherent Excitonic Coupling in an Asymmetric Double InGaAs Quantum Well Arises from Many-Body Effects}
\author{Ga\"el Nardin$^1$, Galan Moody$^{1,2}$, Rohan Singh$^{1,2}$, Travis M. Autry$^{1,2}$, Hebin Li$^1$, Fran\c cois Morier-Genoud$^3$, and Steven T. Cundiff$^{1,2}$}
\affiliation{$^1$ JILA, University of Colorado \& National Institute of Standards and Technology, Boulder CO 80309-0440, USA\\
$^2$ Department of Physics, University of Colorado, Boulder CO 80309-0390, USA\\
$^3$ Laboratory of Quantum Optoelectronics, {\'E}cole Polytechnique F{\'e}d{\'e}rale de Lausanne (EPFL), CH-1015 Lausanne, Switzerland.}

\begin{abstract}
We study an asymmetric double InGaAs quantum well using optical two-dimensional coherent spectroscopy. The collection of zero-quantum, one-quantum and two-quantum two-dimensional spectra provides a unique and comprehensive picture of the double well coherent optical response. Coherent and incoherent contributions to the coupling between the two quantum well excitons are clearly separated. An excellent agreement with density matrix calculations reveals that coherent inter-well coupling originates from many-body interactions.
\end{abstract}

\maketitle


Coupled quantum wells (QWs) are one of the most fundamental topics of quantum mechanics. They can be realized in epitaxially-grown semiconductor materials, where the coupling can be exploited in optoelectronic devices such as quantum cascade lasers \cite{faist_quantum_1994}. Furthermore, since QW and barrier sizes can be tailored, coupled semiconductor QWs can serve as a model for other systems. For example, the absence of vibrational
coupling in semiconductor QWs allows isolation of electronic coupling; this distinctive feature may help understanding extremely efficient energy transfer in light harvesting complexes, where the roles played by electronic and vibrational coupling are under debate \cite{2007Engel_Nature,butkus_vibrational_2012,tiwari_electronic_2013,hayes_engineering_2013}. Semiconductor double QWs (DQWs) have attracted theoretical and experimental attention for more than twenty years. The roles of resonant transfer and wavefunction hybridization \cite{deveaud_subpicosecond_1990,leo_coherent_1991,roskos_coherent_1992}, phonon-assisted tunneling \cite{oberli_optical_1990,roussignol_resonance_1991}, dipole-dipole coupling \cite{batsch_dipole-dipole_1993}, percolation of carriers through imperfect barriers \cite{kim_percolation_1996}, and thermally activated charge transfer \cite{borri_excitation_1999} have been studied and discussed, as well as the formation of indirect excitons \cite{butov_direct_1995,bayer_direct_1996}. However, the role played by many-body effects---which have been shown to dominate the coherent response of semiconductor excitons \cite{li_many-body_2006,turner_persistent_2012}---in the coupling mechanism has been neglected so far.


We use optical two-dimensional coherent spectroscopy (2DCS) to characterize coupling between the QW excitons, which are electron-hole pairs bound together by their Coulomb attraction. 2DCS is an extension of transient four-wave-mixing (FWM) spectroscopy, with the addition of interferometric stabilization of inter-pulse delays and measurement of the signal field. It is an ideal technique to study coupling between resonances, since unfolding one-dimensional spectra onto a second dimension distinguishes quantum beats from polarization interferences \cite{koch_quantum_1992}. Additionally, 2DCS has been demonstrated as a powerful tool for revealing many-body effects in semiconductor nanostructures \cite{li_many-body_2006,turner_persistent_2012}. Several types of 2D spectra---isolating zero-, one-, and two-quantum coherences---have been shown in previous work to reveal information that one-dimensional techniques cannot access \cite{2008Yang_JCP,dai_two-dimensional_2010,cundiff_optical_2012_IEEE,moody_correlation_2013,2008Yang_PRL,2009Stone_Science,karaiskaj_two-quantum_2010,dai_two-dimensional_2012}, but these different types of spectra have never been recorded and analyzed together for a single system so far. Previously, multidimensional spectroscopy showed electronic coherences between excitonic transitions of a GaAs/AlGaAs DQW \cite{li_investigation_2009,davis_three-dimensional_2011,hall_three-dimensional_2013}. However, the presence of heavy and light holes in each GaAs/AlGaAs well gives rise to a plethora of excitonic transitions, thus inhibiting the ability to establish the roles played by different coupling mechanisms.


In this letter, we circumvent this issue by studying strained InGaAs/GaAs QWs. In these structures, the light-hole exciton is not confined in the QW \cite{marzin_optical_1985,moran_nature_1998}. Moreover, the use of InGaAs/GaAs QWs rules out carrier percolation through the barrier as a coupling mechanism \cite{kim_percolation_1996}. InGaAs/GaAs QWs thus provide a semiconductor realization that is as close as possible to the fundamental double well problem of quantum mechanics. We present a set of 2D spectra : zero-, one-, and two-quantum spectra. Each 2D spectrum reveals coupling terms that one-dimensional techniques cannot access, and the collection of them provides a comprehensive picture of the different contributions to the inter-well coupling. The modeling of this unique data set using density matrix calculations shows that the inclusion of many-body effects is essential to reproduce the signatures of the coherent inter-well coupling.



\begin{figure}[htbp]
	\centering
		\includegraphics[width=8.6cm]{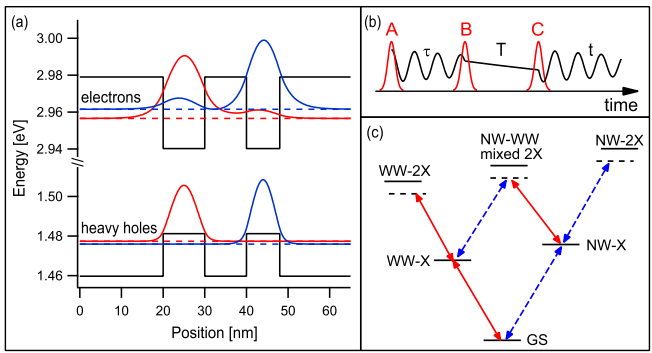}

\caption{(Color online) (a) One dimensional single-particle solutions of the Schr\"odinger equation for the electron and heavy hole for the coupled DQW (8 nm and 10 nm thick InGaAs QWs separated by a GaAs barrier of 10nm thickness). Heavy hole excited states are not shown, since their wavefunctions are essentially orthogonal to electron states.(b) Pulse sequence used in one-quantum and zero-quantum experiments. (c) Energy level diagram used to model the DQW system.}
	\label{Fig1}
\end{figure}

Two samples containing two In$_{0.05}$Ga$_{0.95}$As QWs: an 8 nm thick well (narrow well - NW) and a 10 nm thick well (wide well - WW), are studied. The first sample features a 30 nm thick GaAs barrier between the two QWs. It serves as a reference sample, since no inter-well coupling is expected for this barrier thickness according to one-dimensional Schr\"odinger equation calculations of the electron and hole wavefunctions \footnote{Using \emph{nextnano}, available at www.nextnano.de.}. The second sample features a 10 nm barrier, for which electronic wavefunction coupling between the two QWs is expected. For this barrier size, the hole wavefunctions are localized in their respective QWs, whereas the electron wavefunctions are partially delocalized between the wells, as depicted in the simulations shown in Fig. 1(a). We expect this confinement geometry to allow coherent superpositions of the hybridized wavefunctions to be detected as coherent oscillations in the FWM signal \cite{leo_coherent_1991}. Since the hybridization of wavefunctions is weak, excitons formed from hole and electron wavefunctions mostly confined in the narrow or wide well, respectively, are referred to as the narrow well exciton (NW-X) and wide well exciton (WW-X).


To perform 2DCS experiments, we generate four phase-locked pulses propagating in the box geometry \cite{bristow_versatile_2009}. The pulses are co-circularly polarized, with a duration of $\sim$ 150 fs and center photon energy of 1475 meV obtained from a 76 MHz repetition rate mode-locked laser. Three of the pulses, A, B, and C, with wave vectors $\vec{k}_A$, $\vec{k}_B$, and $\vec{k}_C$, respectively, are incident on the sample, creating an excitation spot of $\sim$50 $\mu$m diameter. The incident photon density is $\sim 2.7\times 10^{11}$ cm$^{-2}$ per pulse or lower, and the coherent response was verified to be in the $\chi^{(3)}$ regime by a power dependence measurement. The sample is kept at a temperature of 15 K in a cold finger cryostat. We detect the FWM signal in the direction $\vec{k}_{FWM}=-\vec{k}_A+\vec{k}_B+\vec{k}_C$, which determines that pulse A acts as a conjugate pulse. The signal is heterodyned with a reference pulse and their interference is spectrally-resolved (spectral resolution: 17 $\mu eV$). Fig.~\ref{Fig1}(b) shows the pulse ordering and time delays ($\tau$, $t$, $T$) for one-quantum and zero-quantum 2DCS. In a one-quantum (zero-quantum) 2D experiment, interferograms are recorded while stepping the delay $\tau$ ($T$). The signal is Fourier transformed with respect to this delay to produce a one-quantum (zero-quantum) 2D spectrum that correlates the excitation (mixing) and emission energies \cite{2008Yang_JCP,dai_two-dimensional_2010,cundiff_optical_2012_IEEE,moody_correlation_2013}. We can also obtain information on two-quantum coherences using a pulse sequence with the conjugated pulse A incident on the sample last, providing a 2D spectrum that correlates emission and two-quantum excitation energies \cite{2008Yang_PRL,2009Stone_Science,karaiskaj_two-quantum_2010,dai_two-dimensional_2012}.



\begin{figure}[t]
	\centering
		\includegraphics[width=8.5cm]{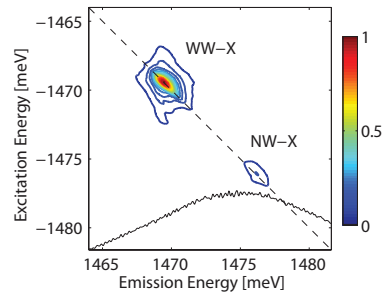}
		\caption{(Color online) Normalized absolute value of a one-quantum 2D spectrum measured on the reference sample (30 nm inter-well barrier), where no inter-well coupling is expected. Solid line: laser spectrum, in arbitrary units.}
		\label{Fig2}
\end{figure}

Figure \ref{Fig2} shows the absolute value of the one-quantum spectrum obtained for the reference sample. The delay between pulses B and C is $T$ = 200 fs. The laser spectrum is presented by the solid line in Fig.~\ref{Fig2}. The 2D spectrum is plotted as a function of the excitation photon energy $\hbar \omega_{\tau}$ and the emission photon energy $\hbar \omega_t$. Two peaks are visible in Fig.~\ref{Fig2}. Located on the diagonal, they correspond to the WW-X and NW-X. The lineshapes of the 2D peaks are typical of inhomogeneously broadened resonances \cite{siemens_resonance_2010}. The absence of cross peaks between the NW-X and WW-X is indicative of an uncoupled system.


\begin{figure}[htbp]
	\centering
		\includegraphics[width=8.5cm]{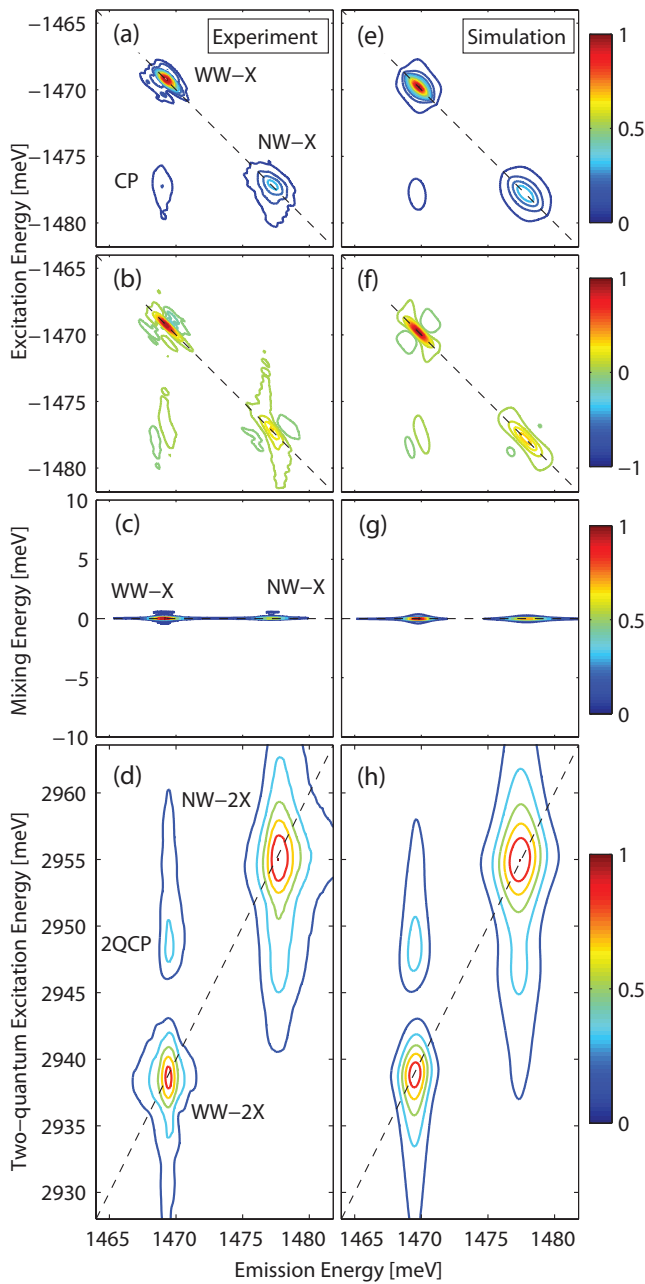}
		\caption{(Color online) Normalized 2D spectra of the asymmetric DQW (10 nm inter-well barrier). (a)-(d)Experimental data. (e)-(h) Density matrix simulations. (a)\&(e) Absolute value of one-quantum spectrum. (b)\&(f) Real part of one-quantum spectrum. (c)\&(g) Absolute value of zero-quantum spectrum. (d)\&(h) Absolute value of two-quantum spectrum.}
		\label{Fig3}
\end{figure}

\begin{figure}[htbp]
	\centering
		\includegraphics[width=8.5cm]{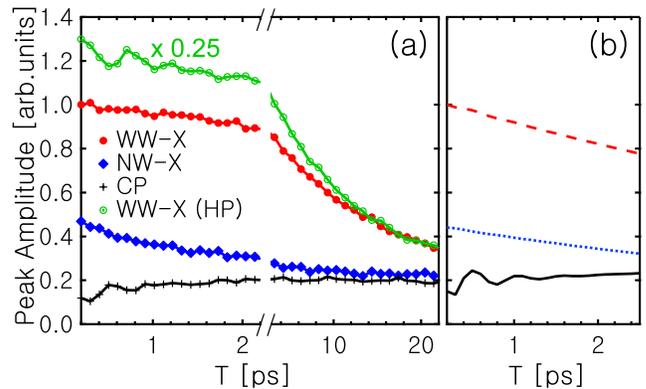}
		\caption{(Color online) Peak amplitudes extracted from (a) experimental, and (b) simulated one-quantum 2D spectra recorded for various values of the delay $T$ between pulses B and C. At longer T (not shown), the simulations predict a monotonic decay of the peak amplitudes. Plain circles (dashed line): WW-X peak amplitude. Squares (dotted line): NW-X peak amplitude. Crosses (solid line): CP amplitude. Empty circles: WW-X peak amplitude for a higher excitation power.}
		\label{Fig4}
\end{figure}

We show the one-quantum 2D spectrum obtained from the sample having a 10 nm inter-well barrier in Fig.~\ref{Fig3}(a), taken with the same experimental conditions as for the reference sample spectrum in Fig.~\ref{Fig2}. We observe two diagonal peaks, corresponding to the excitonic transitions between electron and holes mostly confined in their respective wells: NW-X and WW-X. We do not observe additional indirect excitonic transitions, formed by electrons and holes mostly confined in separated wells, which have a weak oscillator strength due to the low value of the overlap integral between their electron and hole wavefunctions. The energy spacing between WW-X and NW-X is larger than for the reference sample by $\sim$1.1 meV. This number is consistent with single particle Schr\"odinger equation calculations that predict an energy splitting of 1.4 meV due to weak hybridization of the electronic wavefunctions.

The most interesting feature of this 2D spectrum is the cross peak (CP) appearing at the absorption energy of the NW-X and emission energy of the WW-X. This peak is an unambiguous signature of inter-well excitonic coupling. For this barrier size, the inter-well excitonic energy separation coincides with the heavy hole binding energy in InGaAs QWs \cite{hou_exciton_1990,haines_exciton-binding-energy_1991,Suppl_Materials}. We rule out the possibility of this CP being due to coherence between the excitons and free electron-hole pairs in the WW, considering that (1) no CP is observed for the reference sample in similar excitation conditions, and (2) the CP has a well defined peak energy, contrary to the elongated features characteristic of coherences with the continuum of free electron-hole pairs \cite{2005Borca_CPL,Suppl_Materials}.

In Fig.~\ref{Fig3}(b), we plot the real part of the data, revealing the complex lineshape of the excitonic resonances, which are particularly sensitive to many-body effects, namely Excitation Induced Dephasing (EID) and Excitation Induced Shift (EIS) \cite{li_many-body_2006}. In particular, the clear dispersive lineshape of the CP is a strong indication that EIS plays a key role in the inter-well coupling mechanism.


To investigate the inter-well coupling dynamics, we measure one-quantum 2D spectra for various mixing times $T$ \cite{moody_exciton_2011}. The amplitudes of the diagonal and cross peaks are plotted as a function of $T$ in Fig.~\ref{Fig4}(a). While the diagonal peaks mostly undergo a monotonic decay, the CP starts from a non-zero amplitude at T = 200 fs and increases during the first $\sim$5 picoseconds. During the first picosecond of the CP rise, there are damped oscillations with a period of $\sim$0.5 ps, corresponding to a frequency that matches the energy difference of 8 meV between WW-X and NW-X. Thus, two contributions to the inter-well coupling are evident. (1) A fast (decaying within a few picoseconds), coherent contribution, originating from a non-radiative coherence between the WW-X and NW-X states, similar to what was observed in Ref. \cite{leo_coherent_1991}. (2) A slower, incoherent contribution, due to a population relaxation from NW-X to WW-X. Possible mechanisms for the incoherent population relaxation are scattering with an acoustic phonon or resonant interaction of NW-X with the electron-hole continuum of the WW \cite{batsch_dipole-dipole_1993}.


To probe coherences between excited states, we record a zero-quantum 2D spectrum \cite{2008Yang_JCP,dai_two-dimensional_2010,moody_correlation_2013}. This 2D spectrum is shown in Fig.~\ref{Fig3}(c), plotted as a function of mixing energy $\hbar\omega_T$ and emission energy $\hbar\omega_t$. Two peaks appear at zero mixing energy and at the emission energy of the WW-X and NW-X, corresponding to the system being in a ground or excited population state during the delay time $T$. In a zero-quantum spectrum, coherences between excited states would appear as peaks at non-zero mixing energy. Since no such mixing peak is visible, the zero-quantum spectrum does not provide conclusive evidence that the coupling mechanism giving rise to the  CP in Fig. 3(a) is coherent.


However, a third type of 2D spectrum confirms the coherent nature of the NW-WW coupling. A two-quantum spectrum, as shown in Fig.~\ref{Fig3}(d) enables access to signals associated with two-quantum coherences \cite{2008Yang_PRL,2009Stone_Science,karaiskaj_two-quantum_2010,dai_two-dimensional_2012}, which have been shown to arise from many-body interactions \cite{ferrio_observation_1996}. The two peaks on the diagonal of Fig.~\ref{Fig3}(d) arise from coherent interactions between two excitons in the same QW, whereas the two-quantum cross peak (2QCP) at the NW-X - WW-X two-quantum energy and emitting at the WW energy necessarily requires coherent interactions between the NW-X and WW-X.


To determine the role played by many-body effects, we simulate the nonlinear response by analytically solving a perturbative expansion of the density matrix for a 6-level “tree-system” \cite{Suppl_Materials}, shown in Fig.~\ref{Fig1}(c). This energy level scheme consists of a single ground state (GS), two states corresponding to a single exciton in either the WW-X or NW-X, two doubly-excited states representing the WW and NW two-excitons (WW-2X and NW-2X), and a doubly-excited state representing a mixed NW-WW two-exciton (NW-WW-2X). Many-body interactions are introduced by breaking the symmetry between transitions from the ground state to singly excited states and the transitions from singly to doubly excited states, as suggested in Ref. \cite{bott_influence_1993}. The effect of EIS and EID is modeled by altering the transition energy and dephasing rate of the X $\leftrightarrow$ 2X transitions compared to the GS $\leftrightarrow$ X transitions. The simulated spectra are shown in the right column of Fig.~\ref{Fig3}. They accurately reproduce the experimental features, for a single set of parameters for all simulations. Let us underline that the analysis of any subset of spectra would not be conclusive; only the full collection of data provides enough constraints to reveal the dominant mechanisms responsible for the DQW coherent response \cite{Suppl_Materials}.

The simulations show that the inclusion of many-body effects is essential to model the DQW coherent response in general and the signatures of the coherent inter-well coupling in particular. Without many-body effects, there would not be any off-diagonal peak in Figs.~\ref{Fig3}(e) and (f), and no two-quantum signal at all \footnote{Other model systems, such as a 6-level system without many-body interactions, a 3 level-system, or a single particle model system including coupling via indirect excitonic transitions, fail to predict any two-quantum coherence.}. More specifically, the simulations reveal that the 2QCP in Fig.~\ref{Fig3}(h) arises from an EIS of 0.15 meV of the NW-WW-2X state, and this energy shift is also the origin of the CP in Fig.~\ref{Fig3}(e) and (f). The dispersive lineshape of the CP is also a direct consequence of EIS: in an uncoupled system, the GS $\leftrightarrow$ WW-X and NW-X $\leftrightarrow$ NW-WW-2X transitions would be equivalent but with opposite sign, leading to cancellation of the CP signal; with the asymmetry induced by EIS, they no longer coincide on the emission energy axis, and give rise to the CP and its characteristic dispersive lineshape. Hybridization of the electronic states alone cannot account for the interaction peaks: hybridization results in an energy shift between the excitonic eigenstates, which does not result in the aforementioned symmetry breaking, and thus cannot produce the coupling peaks. It is also found that EID on the NW-X $\leftrightarrow$ NW-2X and WW-X $\leftrightarrow$ WW-2X transitions is the dominant mechanism responsible for the occurrence of the WW-2X and NW-2X diagonal peaks in the two-quantum spectrum of Fig.~\ref{Fig3}(h) \cite{Suppl_Materials}. The absence of an above-diagonal cross peak in Fig.~\ref{Fig3}(a) is also reproduced by the simulations when EID provides a larger dephasing for the NW-WW-2X $\leftrightarrow$ NW-X transition than for the WW-X $\leftrightarrow$ GS. At longer $T$, the dominance of the below-diagonal CP over the above diagonal one is further enhanced due to the incoherent NW-X $\rightarrow$ WW-X population relaxation, incorporated into the model by introducing an additional decay of the NW-X population that becomes a source term for the CP. The simulation of the $T$ dependence of the peak amplitudes in the one-quantum spectrum is shown in Fig.~\ref{Fig4}(b), exhibiting the damped oscillations of the CP, consistently with experimental findings. The simulation also shows that the fast decay of the NW-X - WW-X coherence (and the corresponding fast damping of oscillations in the $T$ dependence) prevents the feature from appearing as a peak in the zero-quantum spectrum of Fig.~\ref{Fig3}(g) because it is too weak with respect to the dominant population terms.

It is interesting to note that at very low power the diagonal peak amplitude does not oscillate as a function of $T$ (red curve in Fig.~\ref{Fig4}(a)). This observation is significant because oscillations in the diagonal peak amplitude as function of $T$ have been taken as evidence for quantum coherence playing a role in photosynthesis \cite{2007Engel_Nature}. If we increase the excitation power, we do observe oscillations in the diagonal peak, as shown by the green curve [WW-X (HP)] in Fig.~\ref{Fig4}(a). We ascribe these oscillations to interference between the third-order and fifth-order contributions to the optical response of the sample.

In conclusion, the study of an asymmetric double InGaAs QW using 2DCS has provided a comprehensive picture of the inter-well coupling mechanisms. 2DCS allows for the clear separation of two contributions to the coupling between the NW and WW heavy hole excitons: a fast coherent coupling between NW-X and WW-X, and a slower incoherent relaxation from NW-X to WW-X. Modeling the coupled QW system using density matrix calculations for a 6-level system including many-body effects reproduces the collection of experimental data for a single set of parameters. Many-body effects, which had been mostly neglected in previous work on coupled QWs, are shown to be essential to reproduce the DQW coherent response. In particular, it is found that the coherent inter-well coupling is a direct consequence of many-body interactions between NW-X and WW-X. This result is of interest for quantum cascade lasers, whose optical properties have been predicted to be affected by many-body interactions \cite{liu_microscopic_2012}.This work also highlights the potential of InGaAs QWs as a model system to study interacting excitons: the absence of light hole excitons, charge percolation through the QW barrier, and vibrational coupling provides a cleaner picture of the inter-well coupling. Therefore, it may provide useful insights into the coupling mechanisms that are responsible for energy transfer between light harvesting complexes \cite{butkus_vibrational_2012,tiwari_electronic_2013}.

The work at JILA was primarily supported by the Chemical Sciences, Geosciences, and Energy Biosciences Division,
Office of Basic Energy Science, Office of Science, U.S. Department of Energy under Award\# DE-FG02-02ER15346, as well NIST. G.N. acknowledges support by the Swiss National Science Foundation (SNSF).





\begin{thebibliography}{43}%
\makeatletter
\providecommand \@ifxundefined [1]{%
 \@ifx{#1\undefined}
}%
\providecommand \@ifnum [1]{%
 \ifnum #1\expandafter \@firstoftwo
 \else \expandafter \@secondoftwo
 \fi
}%
\providecommand \@ifx [1]{%
 \ifx #1\expandafter \@firstoftwo
 \else \expandafter \@secondoftwo
 \fi
}%
\providecommand \natexlab [1]{#1}%
\providecommand \enquote  [1]{``#1''}%
\providecommand \bibnamefont  [1]{#1}%
\providecommand \bibfnamefont [1]{#1}%
\providecommand \citenamefont [1]{#1}%
\providecommand \href@noop [0]{\@secondoftwo}%
\providecommand \href [0]{\begingroup \@sanitize@url \@href}%
\providecommand \@href[1]{\@@startlink{#1}\@@href}%
\providecommand \@@href[1]{\endgroup#1\@@endlink}%
\providecommand \@sanitize@url [0]{\catcode `\\12\catcode `\$12\catcode
  `\&12\catcode `\#12\catcode `\^12\catcode `\_12\catcode `\%12\relax}%
\providecommand \@@startlink[1]{}%
\providecommand \@@endlink[0]{}%
\providecommand \url  [0]{\begingroup\@sanitize@url \@url }%
\providecommand \@url [1]{\endgroup\@href {#1}{\urlprefix }}%
\providecommand \urlprefix  [0]{URL }%
\providecommand \Eprint [0]{\href }%
\providecommand \doibase [0]{http://dx.doi.org/}%
\providecommand \selectlanguage [0]{\@gobble}%
\providecommand \bibinfo  [0]{\@secondoftwo}%
\providecommand \bibfield  [0]{\@secondoftwo}%
\providecommand \translation [1]{[#1]}%
\providecommand \BibitemOpen [0]{}%
\providecommand \bibitemStop [0]{}%
\providecommand \bibitemNoStop [0]{.\EOS\space}%
\providecommand \EOS [0]{\spacefactor3000\relax}%
\providecommand \BibitemShut  [1]{\csname bibitem#1\endcsname}%
\let\auto@bib@innerbib\@empty
\bibitem [{\citenamefont {Faist}\ \emph {et~al.}(1994)\citenamefont {Faist},
  \citenamefont {Capasso}, \citenamefont {Sivco}, \citenamefont {Sirtori},
  \citenamefont {Hutchinson},\ and\ \citenamefont {Cho}}]{faist_quantum_1994}%
  \BibitemOpen
  \bibfield  {author} {\bibinfo {author} {\bibfnamefont {J.}~\bibnamefont
  {Faist}}, \bibinfo {author} {\bibfnamefont {F.}~\bibnamefont {Capasso}},
  \bibinfo {author} {\bibfnamefont {D.~L.}\ \bibnamefont {Sivco}}, \bibinfo
  {author} {\bibfnamefont {C.}~\bibnamefont {Sirtori}}, \bibinfo {author}
  {\bibfnamefont {A.~L.}\ \bibnamefont {Hutchinson}}, \ and\ \bibinfo {author}
  {\bibfnamefont {A.~Y.}\ \bibnamefont {Cho}},\ }\href {\doibase
  10.1126/science.264.5158.553} {\bibfield  {journal} {\bibinfo  {journal}
  {Science}\ }\textbf {\bibinfo {volume} {264}},\ \bibinfo {pages} {553}
  (\bibinfo {year} {1994})}\BibitemShut {NoStop}%
\bibitem [{\citenamefont {Engel}\ \emph {et~al.}(2007)\citenamefont {Engel},
  \citenamefont {Calhoun}, \citenamefont {Read}, \citenamefont {Ahn},
  \citenamefont {Mancal}, \citenamefont {Cheng}, \citenamefont {Blankenship},\
  and\ \citenamefont {Fleming}}]{2007Engel_Nature}%
  \BibitemOpen
  \bibfield  {author} {\bibinfo {author} {\bibfnamefont {G.~S.}\ \bibnamefont
  {Engel}}, \bibinfo {author} {\bibfnamefont {T.~R.}\ \bibnamefont {Calhoun}},
  \bibinfo {author} {\bibfnamefont {E.~L.}\ \bibnamefont {Read}}, \bibinfo
  {author} {\bibfnamefont {T.-K.}\ \bibnamefont {Ahn}}, \bibinfo {author}
  {\bibfnamefont {T.}~\bibnamefont {Mancal}}, \bibinfo {author} {\bibfnamefont
  {Y.-C.}\ \bibnamefont {Cheng}}, \bibinfo {author} {\bibfnamefont {R.~E.}\
  \bibnamefont {Blankenship}}, \ and\ \bibinfo {author} {\bibfnamefont {G.~R.}\
  \bibnamefont {Fleming}},\ }\href@noop {} {\bibfield  {journal} {\bibinfo
  {journal} {Nature}\ }\textbf {\bibinfo {volume} {446}},\ \bibinfo {pages}
  {782} (\bibinfo {year} {2007})}\BibitemShut {NoStop}%
\bibitem [{\citenamefont {Butkus}\ \emph {et~al.}(2012)\citenamefont {Butkus},
  \citenamefont {Zigmantas}, \citenamefont {Valkunas},\ and\ \citenamefont
  {Abramavicius}}]{butkus_vibrational_2012}%
  \BibitemOpen
  \bibfield  {author} {\bibinfo {author} {\bibfnamefont {V.}~\bibnamefont
  {Butkus}}, \bibinfo {author} {\bibfnamefont {D.}~\bibnamefont {Zigmantas}},
  \bibinfo {author} {\bibfnamefont {L.}~\bibnamefont {Valkunas}}, \ and\
  \bibinfo {author} {\bibfnamefont {D.}~\bibnamefont {Abramavicius}},\ }\href
  {\doibase 10.1016/j.cplett.2012.07.014} {\bibfield  {journal} {\bibinfo
  {journal} {Chemical Physics Letters}\ }\textbf {\bibinfo {volume} {545}},\
  \bibinfo {pages} {40} (\bibinfo {year} {2012})}\BibitemShut {NoStop}%
\bibitem [{\citenamefont {Tiwari}\ \emph {et~al.}(2013)\citenamefont {Tiwari},
  \citenamefont {Peters},\ and\ \citenamefont
  {Jonas}}]{tiwari_electronic_2013}%
  \BibitemOpen
  \bibfield  {author} {\bibinfo {author} {\bibfnamefont {V.}~\bibnamefont
  {Tiwari}}, \bibinfo {author} {\bibfnamefont {W.~K.}\ \bibnamefont {Peters}},
  \ and\ \bibinfo {author} {\bibfnamefont {D.~M.}\ \bibnamefont {Jonas}},\
  }\href {\doibase 10.1073/pnas.1211157110} {\bibfield  {journal} {\bibinfo
  {journal} {{Proc. Natl. Acad. Sci. USA}}\ }\textbf {\bibinfo {volume}
  {110}},\ \bibinfo {pages} {1203} (\bibinfo {year} {2013})}\BibitemShut
  {NoStop}%
\bibitem [{\citenamefont {Hayes}\ \emph {et~al.}(2013)\citenamefont {Hayes},
  \citenamefont {Griffin},\ and\ \citenamefont
  {Engel}}]{hayes_engineering_2013}%
  \BibitemOpen
  \bibfield  {author} {\bibinfo {author} {\bibfnamefont {D.}~\bibnamefont
  {Hayes}}, \bibinfo {author} {\bibfnamefont {G.~B.}\ \bibnamefont {Griffin}},
  \ and\ \bibinfo {author} {\bibfnamefont {G.~S.}\ \bibnamefont {Engel}},\
  }\href {\doibase 10.1126/science.1233828} {\bibfield  {journal} {\bibinfo
  {journal} {Science}\ }\textbf {\bibinfo {volume} {340}},\ \bibinfo {pages}
  {1431} (\bibinfo {year} {2013})}\BibitemShut {NoStop}%
\bibitem [{\citenamefont {Deveaud}\ \emph {et~al.}(1990)\citenamefont
  {Deveaud}, \citenamefont {Chomette}, \citenamefont {Clerot}, \citenamefont
  {Auvray}, \citenamefont {Regreny}, \citenamefont {Ferreira},\ and\
  \citenamefont {Bastard}}]{deveaud_subpicosecond_1990}%
  \BibitemOpen
  \bibfield  {author} {\bibinfo {author} {\bibfnamefont {B.}~\bibnamefont
  {Deveaud}}, \bibinfo {author} {\bibfnamefont {A.}~\bibnamefont {Chomette}},
  \bibinfo {author} {\bibfnamefont {F.}~\bibnamefont {Clerot}}, \bibinfo
  {author} {\bibfnamefont {P.}~\bibnamefont {Auvray}}, \bibinfo {author}
  {\bibfnamefont {A.}~\bibnamefont {Regreny}}, \bibinfo {author} {\bibfnamefont
  {R.}~\bibnamefont {Ferreira}}, \ and\ \bibinfo {author} {\bibfnamefont
  {G.}~\bibnamefont {Bastard}},\ }\href {\doibase 10.1103/PhysRevB.42.7021}
  {\bibfield  {journal} {\bibinfo  {journal} {Phys. Rev. B}\ }\textbf {\bibinfo
  {volume} {42}},\ \bibinfo {pages} {7021} (\bibinfo {year}
  {1990})}\BibitemShut {NoStop}%
\bibitem [{\citenamefont {Leo}\ \emph {et~al.}(1991)\citenamefont {Leo},
  \citenamefont {Shah}, \citenamefont {G\"obel}, \citenamefont {Damen},
  \citenamefont {Schmitt-Rink}, \citenamefont {Sch\"afer},\ and\ \citenamefont
  {K\"ohler}}]{leo_coherent_1991}%
  \BibitemOpen
  \bibfield  {author} {\bibinfo {author} {\bibfnamefont {K.}~\bibnamefont
  {Leo}}, \bibinfo {author} {\bibfnamefont {J.}~\bibnamefont {Shah}}, \bibinfo
  {author} {\bibfnamefont {E.~O.}\ \bibnamefont {G\"obel}}, \bibinfo {author}
  {\bibfnamefont {T.~C.}\ \bibnamefont {Damen}}, \bibinfo {author}
  {\bibfnamefont {S.}~\bibnamefont {Schmitt-Rink}}, \bibinfo {author}
  {\bibfnamefont {W.}~\bibnamefont {Sch\"afer}}, \ and\ \bibinfo {author}
  {\bibfnamefont {K.}~\bibnamefont {K\"ohler}},\ }\href {\doibase
  10.1103/PhysRevLett.66.201} {\bibfield  {journal} {\bibinfo  {journal} {Phys.
  Rev. Lett.}\ }\textbf {\bibinfo {volume} {66}},\ \bibinfo {pages} {201}
  (\bibinfo {year} {1991})}\BibitemShut {NoStop}%
\bibitem [{\citenamefont {Roskos}\ \emph {et~al.}(1992)\citenamefont {Roskos},
  \citenamefont {Nuss}, \citenamefont {Shah}, \citenamefont {Leo},
  \citenamefont {Miller}, \citenamefont {Fox}, \citenamefont {Schmitt-Rink},\
  and\ \citenamefont {K\"ohler}}]{roskos_coherent_1992}%
  \BibitemOpen
  \bibfield  {author} {\bibinfo {author} {\bibfnamefont {H.~G.}\ \bibnamefont
  {Roskos}}, \bibinfo {author} {\bibfnamefont {M.~C.}\ \bibnamefont {Nuss}},
  \bibinfo {author} {\bibfnamefont {J.}~\bibnamefont {Shah}}, \bibinfo {author}
  {\bibfnamefont {K.}~\bibnamefont {Leo}}, \bibinfo {author} {\bibfnamefont
  {D.~A.~B.}\ \bibnamefont {Miller}}, \bibinfo {author} {\bibfnamefont {A.~M.}\
  \bibnamefont {Fox}}, \bibinfo {author} {\bibfnamefont {S.}~\bibnamefont
  {Schmitt-Rink}}, \ and\ \bibinfo {author} {\bibfnamefont {K.}~\bibnamefont
  {K\"ohler}},\ }\href {\doibase 10.1103/PhysRevLett.68.2216} {\bibfield
  {journal} {\bibinfo  {journal} {Phys. Rev. Lett.}\ }\textbf {\bibinfo
  {volume} {68}},\ \bibinfo {pages} {2216} (\bibinfo {year}
  {1992})}\BibitemShut {NoStop}%
\bibitem [{\citenamefont {Oberli}\ \emph {et~al.}(1990)\citenamefont {Oberli},
  \citenamefont {Shah}, \citenamefont {Damen}, \citenamefont {Kuo},
  \citenamefont {Henry}, \citenamefont {Lary},\ and\ \citenamefont
  {Goodnick}}]{oberli_optical_1990}%
  \BibitemOpen
  \bibfield  {author} {\bibinfo {author} {\bibfnamefont {D.~Y.}\ \bibnamefont
  {Oberli}}, \bibinfo {author} {\bibfnamefont {J.}~\bibnamefont {Shah}},
  \bibinfo {author} {\bibfnamefont {T.~C.}\ \bibnamefont {Damen}}, \bibinfo
  {author} {\bibfnamefont {J.~M.}\ \bibnamefont {Kuo}}, \bibinfo {author}
  {\bibfnamefont {J.~E.}\ \bibnamefont {Henry}}, \bibinfo {author}
  {\bibfnamefont {J.}~\bibnamefont {Lary}}, \ and\ \bibinfo {author}
  {\bibfnamefont {S.~M.}\ \bibnamefont {Goodnick}},\ }\href {\doibase
  doi:10.1063/1.102525} {\bibfield  {journal} {\bibinfo  {journal} {Appl. Phys.
  Lett.}\ }\textbf {\bibinfo {volume} {56}},\ \bibinfo {pages} {1239} (\bibinfo
  {year} {1990})}\BibitemShut {NoStop}%
\bibitem [{\citenamefont {Roussignol}\ \emph {et~al.}(1991)\citenamefont
  {Roussignol}, \citenamefont {Vinattieri}, \citenamefont {Carraresi},
  \citenamefont {Colocci},\ and\ \citenamefont
  {Fasolino}}]{roussignol_resonance_1991}%
  \BibitemOpen
  \bibfield  {author} {\bibinfo {author} {\bibfnamefont {P.}~\bibnamefont
  {Roussignol}}, \bibinfo {author} {\bibfnamefont {A.}~\bibnamefont
  {Vinattieri}}, \bibinfo {author} {\bibfnamefont {L.}~\bibnamefont
  {Carraresi}}, \bibinfo {author} {\bibfnamefont {M.}~\bibnamefont {Colocci}},
  \ and\ \bibinfo {author} {\bibfnamefont {A.}~\bibnamefont {Fasolino}},\
  }\href {\doibase 10.1103/PhysRevB.44.8873} {\bibfield  {journal} {\bibinfo
  {journal} {Phys. Rev. B}\ }\textbf {\bibinfo {volume} {44}},\ \bibinfo
  {pages} {8873} (\bibinfo {year} {1991})}\BibitemShut {NoStop}%
\bibitem [{\citenamefont {Batsch}\ \emph {et~al.}(1993)\citenamefont {Batsch},
  \citenamefont {Meier}, \citenamefont {Thomas}, \citenamefont {Lindberg},
  \citenamefont {Koch},\ and\ \citenamefont
  {Shah}}]{batsch_dipole-dipole_1993}%
  \BibitemOpen
  \bibfield  {author} {\bibinfo {author} {\bibfnamefont {M.}~\bibnamefont
  {Batsch}}, \bibinfo {author} {\bibfnamefont {T.}~\bibnamefont {Meier}},
  \bibinfo {author} {\bibfnamefont {P.}~\bibnamefont {Thomas}}, \bibinfo
  {author} {\bibfnamefont {M.}~\bibnamefont {Lindberg}}, \bibinfo {author}
  {\bibfnamefont {S.~W.}\ \bibnamefont {Koch}}, \ and\ \bibinfo {author}
  {\bibfnamefont {J.}~\bibnamefont {Shah}},\ }\href {\doibase
  10.1103/PhysRevB.48.11817} {\bibfield  {journal} {\bibinfo  {journal} {Phys.
  Rev. B}\ }\textbf {\bibinfo {volume} {48}},\ \bibinfo {pages} {11817}
  (\bibinfo {year} {1993})}\BibitemShut {NoStop}%
\bibitem [{\citenamefont {Kim}\ \emph {et~al.}(1996)\citenamefont {Kim},
  \citenamefont {Ko}, \citenamefont {Kim}, \citenamefont {Rhee}, \citenamefont
  {Hohng}, \citenamefont {Yee}, \citenamefont {Kim}, \citenamefont {Woo},
  \citenamefont {Choi}, \citenamefont {Ihm}, \citenamefont {Woo},\ and\
  \citenamefont {Kang}}]{kim_percolation_1996}%
  \BibitemOpen
  \bibfield  {author} {\bibinfo {author} {\bibfnamefont {D.~S.}\ \bibnamefont
  {Kim}}, \bibinfo {author} {\bibfnamefont {H.~S.}\ \bibnamefont {Ko}},
  \bibinfo {author} {\bibfnamefont {Y.~M.}\ \bibnamefont {Kim}}, \bibinfo
  {author} {\bibfnamefont {S.~J.}\ \bibnamefont {Rhee}}, \bibinfo {author}
  {\bibfnamefont {S.~C.}\ \bibnamefont {Hohng}}, \bibinfo {author}
  {\bibfnamefont {Y.~H.}\ \bibnamefont {Yee}}, \bibinfo {author} {\bibfnamefont
  {W.~S.}\ \bibnamefont {Kim}}, \bibinfo {author} {\bibfnamefont {J.~C.}\
  \bibnamefont {Woo}}, \bibinfo {author} {\bibfnamefont {H.~J.}\ \bibnamefont
  {Choi}}, \bibinfo {author} {\bibfnamefont {J.}~\bibnamefont {Ihm}}, \bibinfo
  {author} {\bibfnamefont {D.~H.}\ \bibnamefont {Woo}}, \ and\ \bibinfo
  {author} {\bibfnamefont {K.~N.}\ \bibnamefont {Kang}},\ }\href {\doibase
  10.1103/PhysRevB.54.14580} {\bibfield  {journal} {\bibinfo  {journal} {Phys.
  Rev. B}\ }\textbf {\bibinfo {volume} {54}},\ \bibinfo {pages} {14580}
  (\bibinfo {year} {1996})}\BibitemShut {NoStop}%
\bibitem [{\citenamefont {Borri}\ \emph {et~al.}(1999)\citenamefont {Borri},
  \citenamefont {Colocci}, \citenamefont {Gurioli}, \citenamefont {Patané},
  \citenamefont {Alessi}, \citenamefont {Capizzi},\ and\ \citenamefont
  {Martelli}}]{borri_excitation_1999}%
  \BibitemOpen
  \bibfield  {author} {\bibinfo {author} {\bibfnamefont {P.}~\bibnamefont
  {Borri}}, \bibinfo {author} {\bibfnamefont {M.}~\bibnamefont {Colocci}},
  \bibinfo {author} {\bibfnamefont {M.}~\bibnamefont {Gurioli}}, \bibinfo
  {author} {\bibfnamefont {A.}~\bibnamefont {Patané}}, \bibinfo {author}
  {\bibfnamefont {M.}~\bibnamefont {Alessi}}, \bibinfo {author} {\bibfnamefont
  {M.}~\bibnamefont {Capizzi}}, \ and\ \bibinfo {author} {\bibfnamefont
  {F.}~\bibnamefont {Martelli}},\ }\href {\doibase
  10.1016/S1386-9477(99)00028-4} {\bibfield  {journal} {\bibinfo  {journal}
  {Physica E}\ }\textbf {\bibinfo {volume} {5}},\ \bibinfo {pages} {73}
  (\bibinfo {year} {1999})}\BibitemShut {NoStop}%
\bibitem [{\citenamefont {Butov}\ \emph {et~al.}(1995)\citenamefont {Butov},
  \citenamefont {Zrenner}, \citenamefont {Abstreiter}, \citenamefont
  {Petinova},\ and\ \citenamefont {Eberl}}]{butov_direct_1995}%
  \BibitemOpen
  \bibfield  {author} {\bibinfo {author} {\bibfnamefont {L.~V.}\ \bibnamefont
  {Butov}}, \bibinfo {author} {\bibfnamefont {A.}~\bibnamefont {Zrenner}},
  \bibinfo {author} {\bibfnamefont {G.}~\bibnamefont {Abstreiter}}, \bibinfo
  {author} {\bibfnamefont {A.~V.}\ \bibnamefont {Petinova}}, \ and\ \bibinfo
  {author} {\bibfnamefont {K.}~\bibnamefont {Eberl}},\ }\href {\doibase
  10.1103/PhysRevB.52.12153} {\bibfield  {journal} {\bibinfo  {journal} {Phys.
  Rev. B}\ }\textbf {\bibinfo {volume} {52}},\ \bibinfo {pages} {12153}
  (\bibinfo {year} {1995})}\BibitemShut {NoStop}%
\bibitem [{\citenamefont {Bayer}\ \emph {et~al.}(1996)\citenamefont {Bayer},
  \citenamefont {Timofeev}, \citenamefont {Faller}, \citenamefont {Gutbrod},\
  and\ \citenamefont {Forchel}}]{bayer_direct_1996}%
  \BibitemOpen
  \bibfield  {author} {\bibinfo {author} {\bibfnamefont {M.}~\bibnamefont
  {Bayer}}, \bibinfo {author} {\bibfnamefont {V.~B.}\ \bibnamefont {Timofeev}},
  \bibinfo {author} {\bibfnamefont {F.}~\bibnamefont {Faller}}, \bibinfo
  {author} {\bibfnamefont {T.}~\bibnamefont {Gutbrod}}, \ and\ \bibinfo
  {author} {\bibfnamefont {A.}~\bibnamefont {Forchel}},\ }\href {\doibase
  10.1103/PhysRevB.54.8799} {\bibfield  {journal} {\bibinfo  {journal} {Phys.
  Rev. B}\ }\textbf {\bibinfo {volume} {54}},\ \bibinfo {pages} {8799}
  (\bibinfo {year} {1996})}\BibitemShut {NoStop}%
\bibitem [{\citenamefont {Li}\ \emph {et~al.}(2006)\citenamefont {Li},
  \citenamefont {Zhang}, \citenamefont {Borca},\ and\ \citenamefont
  {Cundiff}}]{li_many-body_2006}%
  \BibitemOpen
  \bibfield  {author} {\bibinfo {author} {\bibfnamefont {X.}~\bibnamefont
  {Li}}, \bibinfo {author} {\bibfnamefont {T.}~\bibnamefont {Zhang}}, \bibinfo
  {author} {\bibfnamefont {C.~N.}\ \bibnamefont {Borca}}, \ and\ \bibinfo
  {author} {\bibfnamefont {S.~T.}\ \bibnamefont {Cundiff}},\ }\href {\doibase
  10.1103/PhysRevLett.96.057406} {\bibfield  {journal} {\bibinfo  {journal}
  {Phys. Rev. Lett.}\ }\textbf {\bibinfo {volume} {96}},\ \bibinfo {pages}
  {057406} (\bibinfo {year} {2006})}\BibitemShut {NoStop}%
\bibitem [{\citenamefont {Turner}\ \emph {et~al.}(2012)\citenamefont {Turner},
  \citenamefont {Wen}, \citenamefont {Arias}, \citenamefont {Nelson},
  \citenamefont {Li}, \citenamefont {Moody}, \citenamefont {Siemens},\ and\
  \citenamefont {Cundiff}}]{turner_persistent_2012}%
  \BibitemOpen
  \bibfield  {author} {\bibinfo {author} {\bibfnamefont {D.~B.}\ \bibnamefont
  {Turner}}, \bibinfo {author} {\bibfnamefont {P.}~\bibnamefont {Wen}},
  \bibinfo {author} {\bibfnamefont {D.~H.}\ \bibnamefont {Arias}}, \bibinfo
  {author} {\bibfnamefont {K.~A.}\ \bibnamefont {Nelson}}, \bibinfo {author}
  {\bibfnamefont {H.}~\bibnamefont {Li}}, \bibinfo {author} {\bibfnamefont
  {G.}~\bibnamefont {Moody}}, \bibinfo {author} {\bibfnamefont {M.~E.}\
  \bibnamefont {Siemens}}, \ and\ \bibinfo {author} {\bibfnamefont {S.~T.}\
  \bibnamefont {Cundiff}},\ }\href {\doibase 10.1103/PhysRevB.85.201303}
  {\bibfield  {journal} {\bibinfo  {journal} {Phys. Rev. B}\ }\textbf {\bibinfo
  {volume} {85}},\ \bibinfo {pages} {201303} (\bibinfo {year}
  {2012})}\BibitemShut {NoStop}%
\bibitem [{\citenamefont {Koch}\ \emph {et~al.}(1992)\citenamefont {Koch},
  \citenamefont {Feldmann}, \citenamefont {von Plessen}, \citenamefont
  {G\"obel}, \citenamefont {Thomas},\ and\ \citenamefont
  {K\"ohler}}]{koch_quantum_1992}%
  \BibitemOpen
  \bibfield  {author} {\bibinfo {author} {\bibfnamefont {M.}~\bibnamefont
  {Koch}}, \bibinfo {author} {\bibfnamefont {J.}~\bibnamefont {Feldmann}},
  \bibinfo {author} {\bibfnamefont {G.}~\bibnamefont {von Plessen}}, \bibinfo
  {author} {\bibfnamefont {E.~O.}\ \bibnamefont {G\"obel}}, \bibinfo {author}
  {\bibfnamefont {P.}~\bibnamefont {Thomas}}, \ and\ \bibinfo {author}
  {\bibfnamefont {K.}~\bibnamefont {K\"ohler}},\ }\href {\doibase
  10.1103/PhysRevLett.69.3631} {\bibfield  {journal} {\bibinfo  {journal}
  {Phys. Rev. Lett.}\ }\textbf {\bibinfo {volume} {69}},\ \bibinfo {pages}
  {3631} (\bibinfo {year} {1992})}\BibitemShut {NoStop}%
\bibitem [{\citenamefont {Yang}\ \emph {et~al.}(2008)\citenamefont {Yang},
  \citenamefont {Zhang}, \citenamefont {Bristow}, \citenamefont {Cundiff},\
  and\ \citenamefont {Mukamel}}]{2008Yang_JCP}%
  \BibitemOpen
  \bibfield  {author} {\bibinfo {author} {\bibfnamefont {L.}~\bibnamefont
  {Yang}}, \bibinfo {author} {\bibfnamefont {T.}~\bibnamefont {Zhang}},
  \bibinfo {author} {\bibfnamefont {A.~D.}\ \bibnamefont {Bristow}}, \bibinfo
  {author} {\bibfnamefont {S.~T.}\ \bibnamefont {Cundiff}}, \ and\ \bibinfo
  {author} {\bibfnamefont {S.}~\bibnamefont {Mukamel}},\ }\href@noop {}
  {\bibfield  {journal} {\bibinfo  {journal} {J. Chem. Phys.}\ }\textbf
  {\bibinfo {volume} {129}},\ \bibinfo {pages} {234711} (\bibinfo {year}
  {2008})}\BibitemShut {NoStop}%
\bibitem [{\citenamefont {Dai}\ \emph {et~al.}(2010)\citenamefont {Dai},
  \citenamefont {Bristow}, \citenamefont {Karaiskaj},\ and\ \citenamefont
  {Cundiff}}]{dai_two-dimensional_2010}%
  \BibitemOpen
  \bibfield  {author} {\bibinfo {author} {\bibfnamefont {X.}~\bibnamefont
  {Dai}}, \bibinfo {author} {\bibfnamefont {A.~D.}\ \bibnamefont {Bristow}},
  \bibinfo {author} {\bibfnamefont {D.}~\bibnamefont {Karaiskaj}}, \ and\
  \bibinfo {author} {\bibfnamefont {S.~T.}\ \bibnamefont {Cundiff}},\ }\href
  {\doibase 10.1103/PhysRevA.82.052503} {\bibfield  {journal} {\bibinfo
  {journal} {Phys. Rev. A}\ }\textbf {\bibinfo {volume} {82}},\ \bibinfo
  {pages} {052503} (\bibinfo {year} {2010})}\BibitemShut {NoStop}%
\bibitem [{\citenamefont {Cundiff}\ \emph {et~al.}(2012)\citenamefont
  {Cundiff}, \citenamefont {Bristow}, \citenamefont {Siemens}, \citenamefont
  {Li}, \citenamefont {Moody}, \citenamefont {Karaiskaj}, \citenamefont {Dai},\
  and\ \citenamefont {Zhang}}]{cundiff_optical_2012_IEEE}%
  \BibitemOpen
  \bibfield  {author} {\bibinfo {author} {\bibfnamefont {S.}~\bibnamefont
  {Cundiff}}, \bibinfo {author} {\bibfnamefont {A.}~\bibnamefont {Bristow}},
  \bibinfo {author} {\bibfnamefont {M.}~\bibnamefont {Siemens}}, \bibinfo
  {author} {\bibfnamefont {H.}~\bibnamefont {Li}}, \bibinfo {author}
  {\bibfnamefont {G.}~\bibnamefont {Moody}}, \bibinfo {author} {\bibfnamefont
  {D.}~\bibnamefont {Karaiskaj}}, \bibinfo {author} {\bibfnamefont
  {X.}~\bibnamefont {Dai}}, \ and\ \bibinfo {author} {\bibfnamefont
  {T.}~\bibnamefont {Zhang}},\ }\href {\doibase 10.1109/JSTQE.2011.2123876}
  {\bibfield  {journal} {\bibinfo  {journal} {{IEEE} J. Sel. Topics Quantum
  Electr}\ }\textbf {\bibinfo {volume} {18}},\ \bibinfo {pages} {318} (\bibinfo
  {year} {2012})}\BibitemShut {NoStop}%
\bibitem [{\citenamefont {Moody}\ \emph {et~al.}(2013)\citenamefont {Moody},
  \citenamefont {Singh}, \citenamefont {Li}, \citenamefont {Akimov},
  \citenamefont {Bayer}, \citenamefont {Reuter}, \citenamefont {Wieck},\ and\
  \citenamefont {Cundiff}}]{moody_correlation_2013}%
  \BibitemOpen
  \bibfield  {author} {\bibinfo {author} {\bibfnamefont {G.}~\bibnamefont
  {Moody}}, \bibinfo {author} {\bibfnamefont {R.}~\bibnamefont {Singh}},
  \bibinfo {author} {\bibfnamefont {H.}~\bibnamefont {Li}}, \bibinfo {author}
  {\bibfnamefont {I.}~\bibnamefont {Akimov}}, \bibinfo {author} {\bibfnamefont
  {M.}~\bibnamefont {Bayer}}, \bibinfo {author} {\bibfnamefont
  {D.}~\bibnamefont {Reuter}}, \bibinfo {author} {\bibfnamefont
  {A.}~\bibnamefont {Wieck}}, \ and\ \bibinfo {author} {\bibfnamefont
  {S.}~\bibnamefont {Cundiff}},\ }\href {\doibase 10.1016/j.ssc.2013.03.025}
  {\bibfield  {journal} {\bibinfo  {journal} {Solid State Commun.}\ }\textbf
  {\bibinfo {volume} {163}},\ \bibinfo {pages} {65} (\bibinfo {year}
  {2013})}\BibitemShut {NoStop}%
\bibitem [{\citenamefont {Yang}\ and\ \citenamefont
  {Mukamel}(2008)}]{2008Yang_PRL}%
  \BibitemOpen
  \bibfield  {author} {\bibinfo {author} {\bibfnamefont {L.}~\bibnamefont
  {Yang}}\ and\ \bibinfo {author} {\bibfnamefont {S.}~\bibnamefont {Mukamel}},\
  }\href@noop {} {\bibfield  {journal} {\bibinfo  {journal} {Phys. Rev. Lett.}\
  }\textbf {\bibinfo {volume} {100}},\ \bibinfo {pages} {057402} (\bibinfo
  {year} {2008})}\BibitemShut {NoStop}%
\bibitem [{\citenamefont {Stone}\ \emph {et~al.}(2009)\citenamefont {Stone},
  \citenamefont {Gundogdu}, \citenamefont {Turner}, \citenamefont {Li},
  \citenamefont {Cundiff},\ and\ \citenamefont {Nelson}}]{2009Stone_Science}%
  \BibitemOpen
  \bibfield  {author} {\bibinfo {author} {\bibfnamefont {K.~W.}\ \bibnamefont
  {Stone}}, \bibinfo {author} {\bibfnamefont {K.}~\bibnamefont {Gundogdu}},
  \bibinfo {author} {\bibfnamefont {D.~B.}\ \bibnamefont {Turner}}, \bibinfo
  {author} {\bibfnamefont {X.}~\bibnamefont {Li}}, \bibinfo {author}
  {\bibfnamefont {S.~T.}\ \bibnamefont {Cundiff}}, \ and\ \bibinfo {author}
  {\bibfnamefont {K.~A.}\ \bibnamefont {Nelson}},\ }\href@noop {} {\bibfield
  {journal} {\bibinfo  {journal} {Science}\ }\textbf {\bibinfo {volume}
  {324}},\ \bibinfo {pages} {1169} (\bibinfo {year} {2009})}\BibitemShut
  {NoStop}%
\bibitem [{\citenamefont {Karaiskaj}\ \emph {et~al.}(2010)\citenamefont
  {Karaiskaj}, \citenamefont {Bristow}, \citenamefont {Yang}, \citenamefont
  {Dai}, \citenamefont {Mirin}, \citenamefont {Mukamel},\ and\ \citenamefont
  {Cundiff}}]{karaiskaj_two-quantum_2010}%
  \BibitemOpen
  \bibfield  {author} {\bibinfo {author} {\bibfnamefont {D.}~\bibnamefont
  {Karaiskaj}}, \bibinfo {author} {\bibfnamefont {A.~D.}\ \bibnamefont
  {Bristow}}, \bibinfo {author} {\bibfnamefont {L.}~\bibnamefont {Yang}},
  \bibinfo {author} {\bibfnamefont {X.}~\bibnamefont {Dai}}, \bibinfo {author}
  {\bibfnamefont {R.~P.}\ \bibnamefont {Mirin}}, \bibinfo {author}
  {\bibfnamefont {S.}~\bibnamefont {Mukamel}}, \ and\ \bibinfo {author}
  {\bibfnamefont {S.~T.}\ \bibnamefont {Cundiff}},\ }\href {\doibase
  10.1103/PhysRevLett.104.117401} {\bibfield  {journal} {\bibinfo  {journal}
  {Phys. Rev. Lett.}\ }\textbf {\bibinfo {volume} {104}},\ \bibinfo {pages}
  {117401} (\bibinfo {year} {2010})}\BibitemShut {NoStop}%
\bibitem [{\citenamefont {Dai}\ \emph {et~al.}(2012)\citenamefont {Dai},
  \citenamefont {Richter}, \citenamefont {Li}, \citenamefont {Bristow},
  \citenamefont {Falvo}, \citenamefont {Mukamel},\ and\ \citenamefont
  {Cundiff}}]{dai_two-dimensional_2012}%
  \BibitemOpen
  \bibfield  {author} {\bibinfo {author} {\bibfnamefont {X.}~\bibnamefont
  {Dai}}, \bibinfo {author} {\bibfnamefont {M.}~\bibnamefont {Richter}},
  \bibinfo {author} {\bibfnamefont {H.}~\bibnamefont {Li}}, \bibinfo {author}
  {\bibfnamefont {A.~D.}\ \bibnamefont {Bristow}}, \bibinfo {author}
  {\bibfnamefont {C.}~\bibnamefont {Falvo}}, \bibinfo {author} {\bibfnamefont
  {S.}~\bibnamefont {Mukamel}}, \ and\ \bibinfo {author} {\bibfnamefont
  {S.~T.}\ \bibnamefont {Cundiff}},\ }\href {\doibase
  10.1103/PhysRevLett.108.193201} {\bibfield  {journal} {\bibinfo  {journal}
  {Phys. Rev. Lett.}\ }\textbf {\bibinfo {volume} {108}},\ \bibinfo {pages}
  {193201} (\bibinfo {year} {2012})}\BibitemShut {NoStop}%
\bibitem [{\citenamefont {Li}\ \emph {et~al.}(2009)\citenamefont {Li},
  \citenamefont {Zhang}, \citenamefont {Mukamel}, \citenamefont {Mirin},\ and\
  \citenamefont {Cundiff}}]{li_investigation_2009}%
  \BibitemOpen
  \bibfield  {author} {\bibinfo {author} {\bibfnamefont {X.}~\bibnamefont
  {Li}}, \bibinfo {author} {\bibfnamefont {T.}~\bibnamefont {Zhang}}, \bibinfo
  {author} {\bibfnamefont {S.}~\bibnamefont {Mukamel}}, \bibinfo {author}
  {\bibfnamefont {R.~P.}\ \bibnamefont {Mirin}}, \ and\ \bibinfo {author}
  {\bibfnamefont {S.~T.}\ \bibnamefont {Cundiff}},\ }\href {\doibase
  10.1016/j.ssc.2008.12.021} {\bibfield  {journal} {\bibinfo  {journal} {Solid
  State Commun.}\ }\textbf {\bibinfo {volume} {149}},\ \bibinfo {pages} {361}
  (\bibinfo {year} {2009})}\BibitemShut {NoStop}%
\bibitem [{\citenamefont {Davis}\ \emph {et~al.}(2011)\citenamefont {Davis},
  \citenamefont {Hall}, \citenamefont {Dao}, \citenamefont {Nugent},
  \citenamefont {Quiney}, \citenamefont {Tan},\ and\ \citenamefont
  {Jagadish}}]{davis_three-dimensional_2011}%
  \BibitemOpen
  \bibfield  {author} {\bibinfo {author} {\bibfnamefont {J.~A.}\ \bibnamefont
  {Davis}}, \bibinfo {author} {\bibfnamefont {C.~R.}\ \bibnamefont {Hall}},
  \bibinfo {author} {\bibfnamefont {L.~V.}\ \bibnamefont {Dao}}, \bibinfo
  {author} {\bibfnamefont {K.~A.}\ \bibnamefont {Nugent}}, \bibinfo {author}
  {\bibfnamefont {H.~M.}\ \bibnamefont {Quiney}}, \bibinfo {author}
  {\bibfnamefont {H.~H.}\ \bibnamefont {Tan}}, \ and\ \bibinfo {author}
  {\bibfnamefont {C.}~\bibnamefont {Jagadish}},\ }\href {\doibase
  doi:10.1063/1.3613679} {\bibfield  {journal} {\bibinfo  {journal} {J. Chem.
  Phys.}\ }\textbf {\bibinfo {volume} {135}},\ \bibinfo {pages} {044510}
  (\bibinfo {year} {2011})}\BibitemShut {NoStop}%
\bibitem [{\citenamefont {Hall}\ \emph {et~al.}(2013)\citenamefont {Hall},
  \citenamefont {Tollerud}, \citenamefont {Quiney},\ and\ \citenamefont
  {Davis}}]{hall_three-dimensional_2013}%
  \BibitemOpen
  \bibfield  {author} {\bibinfo {author} {\bibfnamefont {C.~R.}\ \bibnamefont
  {Hall}}, \bibinfo {author} {\bibfnamefont {J.~O.}\ \bibnamefont {Tollerud}},
  \bibinfo {author} {\bibfnamefont {H.~M.}\ \bibnamefont {Quiney}}, \ and\
  \bibinfo {author} {\bibfnamefont {J.~A.}\ \bibnamefont {Davis}},\ }\href
  {\doibase 10.1088/1367-2630/15/4/045028} {\bibfield  {journal} {\bibinfo
  {journal} {New J. Phys.}\ }\textbf {\bibinfo {volume} {15}},\ \bibinfo
  {pages} {045028} (\bibinfo {year} {2013})}\BibitemShut {NoStop}%
\bibitem [{\citenamefont {Marzin}\ \emph {et~al.}(1985)\citenamefont {Marzin},
  \citenamefont {Charasse},\ and\ \citenamefont
  {Sermage}}]{marzin_optical_1985}%
  \BibitemOpen
  \bibfield  {author} {\bibinfo {author} {\bibfnamefont {J.-Y.}\ \bibnamefont
  {Marzin}}, \bibinfo {author} {\bibfnamefont {M.~N.}\ \bibnamefont
  {Charasse}}, \ and\ \bibinfo {author} {\bibfnamefont {B.}~\bibnamefont
  {Sermage}},\ }\href {\doibase 10.1103/PhysRevB.31.8298} {\bibfield  {journal}
  {\bibinfo  {journal} {Phys. Rev. B}\ }\textbf {\bibinfo {volume} {31}},\
  \bibinfo {pages} {8298} (\bibinfo {year} {1985})}\BibitemShut {NoStop}%
\bibitem [{\citenamefont {Moran}\ \emph {et~al.}(1998)\citenamefont {Moran},
  \citenamefont {Dawson},\ and\ \citenamefont {Moore}}]{moran_nature_1998}%
  \BibitemOpen
  \bibfield  {author} {\bibinfo {author} {\bibfnamefont {M.}~\bibnamefont
  {Moran}}, \bibinfo {author} {\bibfnamefont {P.}~\bibnamefont {Dawson}}, \
  and\ \bibinfo {author} {\bibfnamefont {K.}~\bibnamefont {Moore}},\ }\href
  {\doibase 10.1016/S0038-1098(98)00163-X} {\bibfield  {journal} {\bibinfo
  {journal} {Solid State Commun.}\ }\textbf {\bibinfo {volume} {107}},\
  \bibinfo {pages} {119} (\bibinfo {year} {1998})}\BibitemShut {NoStop}%
\bibitem [{Note1()}]{Note1}%
  \BibitemOpen
  \bibinfo {note} {Using \protect \emph {nextnano}, available at
  www.nextnano.de.}\BibitemShut {Stop}%
\bibitem [{\citenamefont {Bristow}\ \emph {et~al.}(2009)\citenamefont
  {Bristow}, \citenamefont {Karaiskaj}, \citenamefont {Dai}, \citenamefont
  {Zhang}, \citenamefont {Carlsson}, \citenamefont {Hagen}, \citenamefont
  {Jimenez},\ and\ \citenamefont {Cundiff}}]{bristow_versatile_2009}%
  \BibitemOpen
  \bibfield  {author} {\bibinfo {author} {\bibfnamefont {A.~D.}\ \bibnamefont
  {Bristow}}, \bibinfo {author} {\bibfnamefont {D.}~\bibnamefont {Karaiskaj}},
  \bibinfo {author} {\bibfnamefont {X.}~\bibnamefont {Dai}}, \bibinfo {author}
  {\bibfnamefont {T.}~\bibnamefont {Zhang}}, \bibinfo {author} {\bibfnamefont
  {C.}~\bibnamefont {Carlsson}}, \bibinfo {author} {\bibfnamefont {K.~R.}\
  \bibnamefont {Hagen}}, \bibinfo {author} {\bibfnamefont {R.}~\bibnamefont
  {Jimenez}}, \ and\ \bibinfo {author} {\bibfnamefont {S.~T.}\ \bibnamefont
  {Cundiff}},\ }\href {\doibase 10.1063/1.3184103} {\bibfield  {journal}
  {\bibinfo  {journal} {Rev. Sci. Instrum.}\ }\textbf {\bibinfo {volume}
  {80}},\ \bibinfo {pages} {073108} (\bibinfo {year} {2009})}\BibitemShut
  {NoStop}%
\bibitem [{\citenamefont {Siemens}\ \emph {et~al.}(2010)\citenamefont
  {Siemens}, \citenamefont {Moody}, \citenamefont {Li}, \citenamefont
  {Bristow},\ and\ \citenamefont {Cundiff}}]{siemens_resonance_2010}%
  \BibitemOpen
  \bibfield  {author} {\bibinfo {author} {\bibfnamefont {M.~E.}\ \bibnamefont
  {Siemens}}, \bibinfo {author} {\bibfnamefont {G.}~\bibnamefont {Moody}},
  \bibinfo {author} {\bibfnamefont {H.}~\bibnamefont {Li}}, \bibinfo {author}
  {\bibfnamefont {A.~D.}\ \bibnamefont {Bristow}}, \ and\ \bibinfo {author}
  {\bibfnamefont {S.~T.}\ \bibnamefont {Cundiff}},\ }\href {\doibase
  10.1364/OE.18.017699} {\bibfield  {journal} {\bibinfo  {journal} {Opt.
  Express}\ }\textbf {\bibinfo {volume} {18}},\ \bibinfo {pages} {17699}
  (\bibinfo {year} {2010})}\BibitemShut {NoStop}%
\bibitem [{\citenamefont {Hou}\ \emph {et~al.}(1990)\citenamefont {Hou},
  \citenamefont {Segawa}, \citenamefont {Aoyagi}, \citenamefont {Namba},\ and\
  \citenamefont {Zhou}}]{hou_exciton_1990}%
  \BibitemOpen
  \bibfield  {author} {\bibinfo {author} {\bibfnamefont {H.~Q.}\ \bibnamefont
  {Hou}}, \bibinfo {author} {\bibfnamefont {Y.}~\bibnamefont {Segawa}},
  \bibinfo {author} {\bibfnamefont {Y.}~\bibnamefont {Aoyagi}}, \bibinfo
  {author} {\bibfnamefont {S.}~\bibnamefont {Namba}}, \ and\ \bibinfo {author}
  {\bibfnamefont {J.~M.}\ \bibnamefont {Zhou}},\ }\href {\doibase
  10.1103/PhysRevB.42.1284} {\bibfield  {journal} {\bibinfo  {journal} {Phys.
  Rev. B}\ }\textbf {\bibinfo {volume} {42}},\ \bibinfo {pages} {1284}
  (\bibinfo {year} {1990})}\BibitemShut {NoStop}%
\bibitem [{\citenamefont {Haines}\ \emph {et~al.}(1991)\citenamefont {Haines},
  \citenamefont {Ahmed}, \citenamefont {Adams}, \citenamefont {Mitchell},
  \citenamefont {Agool}, \citenamefont {Pidgeon}, \citenamefont {Cavenett},
  \citenamefont {{O'Reilly}}, \citenamefont {Ghiti},\ and\ \citenamefont
  {Emeny}}]{haines_exciton-binding-energy_1991}%
  \BibitemOpen
  \bibfield  {author} {\bibinfo {author} {\bibfnamefont {M.~J. L.~S.}\
  \bibnamefont {Haines}}, \bibinfo {author} {\bibfnamefont {N.}~\bibnamefont
  {Ahmed}}, \bibinfo {author} {\bibfnamefont {S.~J.~A.}\ \bibnamefont {Adams}},
  \bibinfo {author} {\bibfnamefont {K.}~\bibnamefont {Mitchell}}, \bibinfo
  {author} {\bibfnamefont {I.~R.}\ \bibnamefont {Agool}}, \bibinfo {author}
  {\bibfnamefont {C.~R.}\ \bibnamefont {Pidgeon}}, \bibinfo {author}
  {\bibfnamefont {B.~C.}\ \bibnamefont {Cavenett}}, \bibinfo {author}
  {\bibfnamefont {E.~P.}\ \bibnamefont {{O'Reilly}}}, \bibinfo {author}
  {\bibfnamefont {A.}~\bibnamefont {Ghiti}}, \ and\ \bibinfo {author}
  {\bibfnamefont {M.~T.}\ \bibnamefont {Emeny}},\ }\href {\doibase
  10.1103/PhysRevB.43.11944} {\bibfield  {journal} {\bibinfo  {journal} {Phys.
  Rev. B}\ }\textbf {\bibinfo {volume} {43}},\ \bibinfo {pages} {11944}
  (\bibinfo {year} {1991})}\BibitemShut {NoStop}%
\bibitem [{Sup()}]{Suppl_Materials}%
  \BibitemOpen
  \href@noop {} {}\bibinfo {note} {See Supplemental Material.}\BibitemShut
  {Stop}%
\bibitem [{\citenamefont {Borca}\ \emph {et~al.}(2005)\citenamefont {Borca},
  \citenamefont {Zhang}, \citenamefont {Li},\ and\ \citenamefont
  {Cundiff}}]{2005Borca_CPL}%
  \BibitemOpen
  \bibfield  {author} {\bibinfo {author} {\bibfnamefont {C.~N.}\ \bibnamefont
  {Borca}}, \bibinfo {author} {\bibfnamefont {T.~H.}\ \bibnamefont {Zhang}},
  \bibinfo {author} {\bibfnamefont {X.~Q.}\ \bibnamefont {Li}}, \ and\ \bibinfo
  {author} {\bibfnamefont {S.~T.}\ \bibnamefont {Cundiff}},\ }\href@noop {}
  {\bibfield  {journal} {\bibinfo  {journal} {Chem. Phys. Lett.}\ }\textbf
  {\bibinfo {volume} {416}},\ \bibinfo {pages} {311} (\bibinfo {year}
  {2005})}\BibitemShut {NoStop}%
\bibitem [{\citenamefont {Moody}\ \emph {et~al.}(2011)\citenamefont {Moody},
  \citenamefont {Siemens}, \citenamefont {Bristow}, \citenamefont {Dai},
  \citenamefont {Bracker}, \citenamefont {Gammon},\ and\ \citenamefont
  {Cundiff}}]{moody_exciton_2011}%
  \BibitemOpen
  \bibfield  {author} {\bibinfo {author} {\bibfnamefont {G.}~\bibnamefont
  {Moody}}, \bibinfo {author} {\bibfnamefont {M.~E.}\ \bibnamefont {Siemens}},
  \bibinfo {author} {\bibfnamefont {A.~D.}\ \bibnamefont {Bristow}}, \bibinfo
  {author} {\bibfnamefont {X.}~\bibnamefont {Dai}}, \bibinfo {author}
  {\bibfnamefont {A.~S.}\ \bibnamefont {Bracker}}, \bibinfo {author}
  {\bibfnamefont {D.}~\bibnamefont {Gammon}}, \ and\ \bibinfo {author}
  {\bibfnamefont {S.~T.}\ \bibnamefont {Cundiff}},\ }\href {\doibase
  10.1103/PhysRevB.83.245316} {\bibfield  {journal} {\bibinfo  {journal} {Phys.
  Rev. B}\ }\textbf {\bibinfo {volume} {83}},\ \bibinfo {pages} {245316}
  (\bibinfo {year} {2011})}\BibitemShut {NoStop}%
\bibitem [{\citenamefont {Ferrio}\ and\ \citenamefont
  {Steel}(1996)}]{ferrio_observation_1996}%
  \BibitemOpen
  \bibfield  {author} {\bibinfo {author} {\bibfnamefont {K.~B.}\ \bibnamefont
  {Ferrio}}\ and\ \bibinfo {author} {\bibfnamefont {D.~G.}\ \bibnamefont
  {Steel}},\ }\href {\doibase 10.1103/PhysRevB.54.R5231} {\bibfield  {journal}
  {\bibinfo  {journal} {Phys. Rev. B}\ }\textbf {\bibinfo {volume} {54}},\
  \bibinfo {pages} {R5231} (\bibinfo {year} {1996})}\BibitemShut {NoStop}%
\bibitem [{\citenamefont {Bott}\ \emph {et~al.}(1993)\citenamefont {Bott},
  \citenamefont {Heller}, \citenamefont {Bennhardt}, \citenamefont {Cundiff},
  \citenamefont {Thomas}, \citenamefont {Mayer}, \citenamefont {Smith},
  \citenamefont {Eccleston}, \citenamefont {Kuhl},\ and\ \citenamefont
  {Ploog}}]{bott_influence_1993}%
  \BibitemOpen
  \bibfield  {author} {\bibinfo {author} {\bibfnamefont {K.}~\bibnamefont
  {Bott}}, \bibinfo {author} {\bibfnamefont {O.}~\bibnamefont {Heller}},
  \bibinfo {author} {\bibfnamefont {D.}~\bibnamefont {Bennhardt}}, \bibinfo
  {author} {\bibfnamefont {S.~T.}\ \bibnamefont {Cundiff}}, \bibinfo {author}
  {\bibfnamefont {P.}~\bibnamefont {Thomas}}, \bibinfo {author} {\bibfnamefont
  {E.~J.}\ \bibnamefont {Mayer}}, \bibinfo {author} {\bibfnamefont {G.~O.}\
  \bibnamefont {Smith}}, \bibinfo {author} {\bibfnamefont {R.}~\bibnamefont
  {Eccleston}}, \bibinfo {author} {\bibfnamefont {J.}~\bibnamefont {Kuhl}}, \
  and\ \bibinfo {author} {\bibfnamefont {K.}~\bibnamefont {Ploog}},\ }\href
  {\doibase 10.1103/PhysRevB.48.17418} {\bibfield  {journal} {\bibinfo
  {journal} {Phys. Rev. B}\ }\textbf {\bibinfo {volume} {48}},\ \bibinfo
  {pages} {17418} (\bibinfo {year} {1993})}\BibitemShut {NoStop}%
\bibitem [{Note2()}]{Note2}%
  \BibitemOpen
  \bibinfo {note} {Other model systems, such as a 6-level system without
  many-body interactions, a 3 level-system, or a single particle model system
  including coupling via indirect excitonic transitions, fail to predict any
  two-quantum coherence.}\BibitemShut {Stop}%
\bibitem [{\citenamefont {Liu}\ \emph {et~al.}(2012)\citenamefont {Liu},
  \citenamefont {Lee},\ and\ \citenamefont {Wang}}]{liu_microscopic_2012}%
  \BibitemOpen
  \bibfield  {author} {\bibinfo {author} {\bibfnamefont {T.}~\bibnamefont
  {Liu}}, \bibinfo {author} {\bibfnamefont {K.~E.}\ \bibnamefont {Lee}}, \ and\
  \bibinfo {author} {\bibfnamefont {Q.~J.}\ \bibnamefont {Wang}},\ }\href
  {\doibase 10.1103/PhysRevB.86.235306} {\bibfield  {journal} {\bibinfo
  {journal} {Phys. Rev. B}\ }\textbf {\bibinfo {volume} {86}},\ \bibinfo
  {pages} {235306} (\bibinfo {year} {2012})}\BibitemShut {NoStop}%
\end{thebibliography}

%

\end{document}